# Extracting Three Dimensional Surface Model of Human Kidney from the Visible Human Data Set using Free Software

Kirana Kumara P[*]

*Centre for Product Design and Manufacturing*
*Indian Institute of Science*
*Bangalore, INDIA - 560012*
E-mail: kiranakumarap@gmail.com

[*]Corresponding author: Phone: +91 80 2293 2284; Fax: +91 80 2360 1975

**Abstract**

Three dimensional digital model of a representative human kidney is needed for a surgical simulator that is capable of simulating a laparoscopic surgery involving kidney. Buying a three dimensional computer model of a representative human kidney, or reconstructing a human kidney from an image sequence using commercial software, both involve (sometimes significant amount of) money. In this paper, author has shown that one can obtain a three dimensional surface model of human kidney by making use of images from the Visible Human Data Set and a few free software packages (ImageJ, ITK-SNAP, and MeshLab in particular). Images from the Visible Human Data Set, and the software packages used here, both do not cost anything. Hence, the practice of extracting the geometry of a representative human kidney for free, as illustrated in the present work, could be a free alternative to the use of expensive commercial software or to the purchase of a digital model.

**Introduction**

Laparoscopic surgery is often a substitute for a traditional open surgery, when human kidney is the organ that is to be operated upon [1]; choosing laparoscopic surgery over an open surgery reduces trauma and shortens the recovery time for the patient [1]. But since laparoscopic surgery needs highly skilled surgeons, it is preferable to use a surgical simulator for training and evaluating surgeons [2]. A surgical simulator that can simulate a laparoscopic surgery over a human kidney needs a virtual kidney, i.e., a computer (digital) three dimensional (3D) model of a representative human kidney.

Currently, mainly two approaches are being practiced to obtain the geometry of a representative human kidney. The first approach is to buy a readily available 3D model of a human kidney from an online store (e.g., [3]). The second approach is to use commercial software packages (such as 3D-DOCTOR [4], Mimics [5], Amira [6]) to reconstruct a 3D geometry of a kidney from a two dimensional (2D) image sequence. One can see that both of these approaches cost (sometimes significant amount of) money.

Present work shows that it is possible to obtain a three dimensional surface model of a representative human kidney, completely for free. Present approach is to make use of a few free software packages to extract 3D geometry of human kidney from CT-scan images from the Visible Human Data Set (VHD) (also known as The Visible Human Project Image Data Set, or The Visible Human Project Data Sets) [7]. The free software packages used are ImageJ [8-10], ITK-SNAP [11, 12], and MeshLab [13, 14]. One can note that images from VHD may be downloaded for free, after obtaining a free license [15] from National Library of Medicine (NLM) [16] which is a part of National Institutes of Health (NIH) [17]; VHD is a part of the more ambitious Visible Human Project (VHP) [18].

Approaches similar to the approach presented in the present paper may be found in the present author's previous works [19, 20]. Although the free software packages used in those works are the same as the ones used in the present work, the images were not from VHD; images used in [19] and [20] were downloaded from [21] (the images are no longer accessible as of now; but images were downloadable sometime back). Also, [19] and [20] discuss the reconstruction of a pig liver, while the present work deals with the reconstruction of a human kidney.

Upon conducting literature review, one can see that there are authors who have used VHD together with commercial software packages, e.g., [22]. Also, there are authors who have used images from sources other than VHD and have used commercial software packages to perform 3D reconstruction of biological organs [23-26]. Also, there are authors who have used free software packages to extract geometry of biological organs [27]. But the present author could not find any source in the literature where the three free software packages –

ImageJ, ITK-SNAP and MeshLab – were used to obtain 3D surface model of human kidney from CT-scan images from the VHD.

The practice of extracting the geometry of a representative human kidney for free, as presented in the present work, could be a free alternative to the use of expensive commercial software or to the purchase of a digital model.

## Material and Method

For the present work, CT-scan images from VHD and the three free and open source software packages – ImageJ, ITK-SNAP, and MeshLab – form the material. As far as method is concerned, the three software packages are used to reconstruct 3D models of human kidney from the CT-scan images from the VHD.

VHD contains CT, MRI and cryosection images. In this work, only normal CT-scan images of visible human male and female are used. Present work uses images in the 'png' format since this is the format recommended by VHP. File size of CT-scan images is small, and the images are good enough for reconstructing a 3D model of whole kidney (i.e., inner (or finer) details of kidney are not present; reconstructed 3D model represents just the outer surface of kidney). One can easily identify human kidney in the CT-scan images of VHD.

In the present work, ImageJ 1.42q is used to form an image stack which contains kidney. ITK-SNAP Version 2.0.0 is used for segmentation and 3D reconstruction to the correct scale. MeshLab v1.2.2 is used to control the level of detail in the reconstructed 3D model; it also serves as a tool to smoothen the 3D model and reduce its file size.

Now the method is explained in a bit detail in the following subsections.

### *Using ImageJ to Form an Image Stack*

CT-scan images for visible human male and female are available from head to toe. Out of these images, one has to identify the images which belong to kidney. Upon viewing individual images in ImageJ, and upon consulting [28] and [29], one can conclude that for visible human male, both left are right kidneys are contained between the images 'cvm1551f.png' and 'cvm1692f.png' (47 images in total). Similarly, for visible human female, both left and right kidneys are contained between the images 'cvf1564f.png' and 'cvf1693f.png' (130 images in total). Now these 47 images for male, and 130 images for female, have to be copied into two separate empty folders. Now, to form an image stack for male, select the menu item 'File -> Import -> Image Sequence…', browse to the location of the folder containing 47 images and select the first image in the folder and follow the prompts (with default options); all 47 images are now displayed in ImageJ as an image stack; now, select the menu item 'File -> Save As -> Raw Data…' to save the image stack in the '*.raw' format (where * is the name given). Similar procedure may be followed to obtain an image stack for the female.

*Using ITK-SNAP to Perform Segmentation and 3D Reconstruction*

ITK-SNAP does the segmentation and 3D reconstruction to the correct scale. Hence header information for the images in the image stack is essential. VHD contains header information for each of the images in its database. Upon going through the header files of each of the 47 images of male, one can note that the following header information is identical for all the 47 images: Image matrix size – X = 512, Image matrix size – Y = 512, Image dimension – X = 460 mm, Image dimension –Y = 460 mm, Image pixel size – X = 0.898438, Image pixel size – Y = 0.898438, Screen Format = 16 bit, Spacing between scans = 3 mm. Similarly, the following header information is identical for all the 130 female images: Image matrix size – X = 512, Image matrix size – Y = 512, Image dimension – X = 480 mm, Image dimension –Y = 480 mm, Image pixel size – X = 0.9375, Image pixel size – Y = 0.9375, Screen Format = 16 bit, Spacing between scans = 1 mm.

Now, the method of reconstructing the left kidney of the male is explained in detail, with illustrations. The same method may be employed to reconstruct the right kidney of the male, and the left and right kidneys of the female.

Select the menu item 'File -> Open Greyscale Image…', browse to the location of the image stack for male, follow the prompts and supply the header information. As noted in the first paragraph of this subsection, the 'missing header information' to be supplied for the image stack (for male) is: Image dimensions: X: 512, Y: 512, Z: 47, Voxel spacing: X = 0.898438, Y = 0.898438, Z = 3, Voxel representation: 16 bit, unsigned. Once the header information is supplied, image stack is displayed in ITK-SNAP. One can browse through all the 47 images in the image stack. For illustration purposes, 17th image and 33rd image in the image stack (i.e., 'cvm1602f.png' and 'cvm1650f.png' in the VHD) are shown in Figure 1 and Figure 2 respectively; also, the left and right kidneys are identified in Figure 1 and Figure

2, by making use of illustrations from [28] and [29].

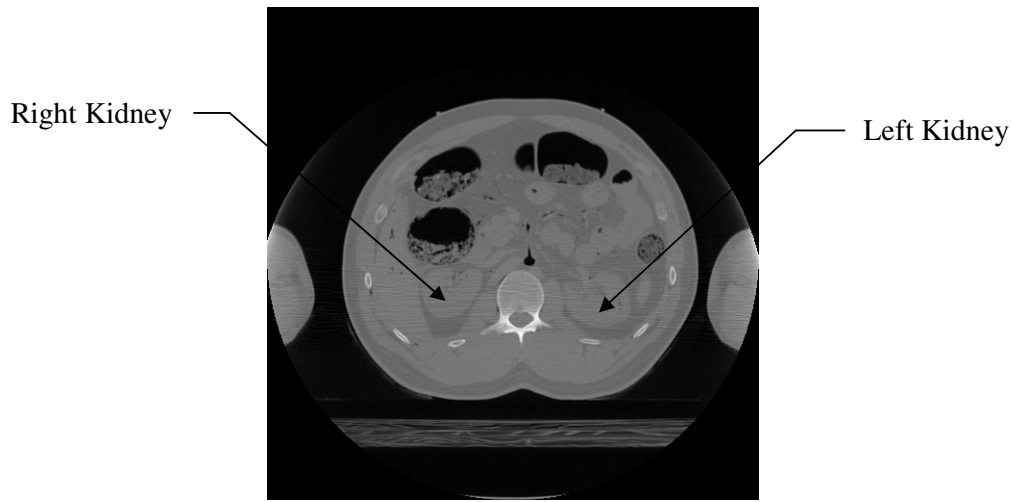

*Figure 1.* The 17th image in the image stack

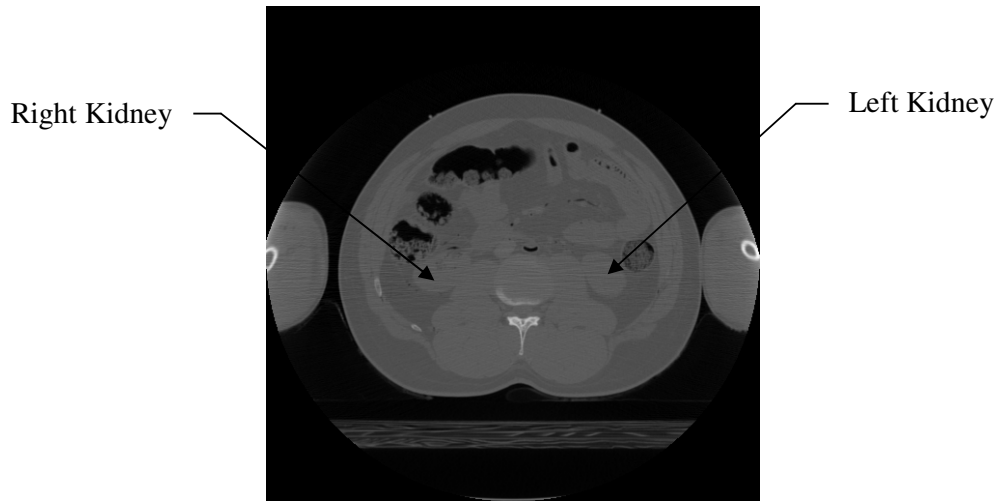

*Figure 2.* The 33rd image in the image stack

Now the task is to do the segmentation. Select 'Polygon tool' from the 'IRIS Toolbox', for slice-by-slice manual segmentation. Select 'continuous' radio button under 'Polygon Tool'. Now click and drag the mouse cursor along the edge of the left kidney (as seen in the axial view (window)), carefully. This draws the contour of the edge of the left kidney. Now, right click on the image, and select the 'accept' button to create the segmentation for the image on display. This process has to be repeated for all images in the image stack, which contain pixels that belong to the left kidney. For illustration purposes, 17th image and 33rd image in the image stack, after segmentation, are shown in Figure 3 and

Figure 4 respectively.

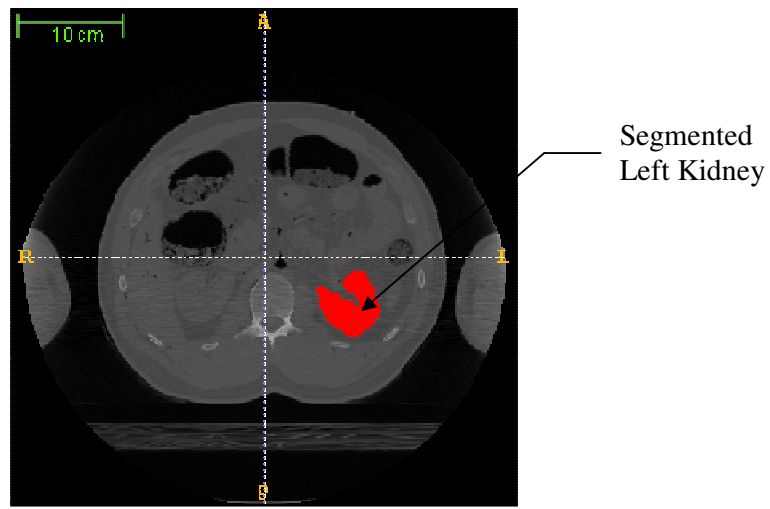

*Figure 3.* *The 17th image in the image stack (after segmentation)*

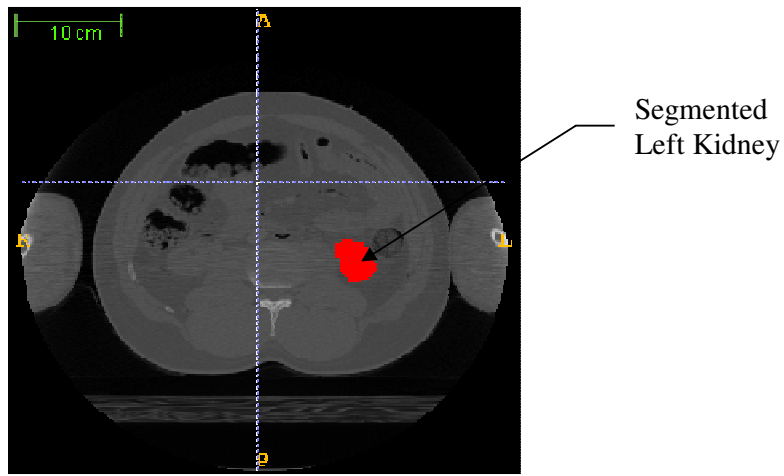

*Figure 4.* *The 33rd image in the image stack (after segmentation)*

Once the segmentation is over, 3D reconstruction is to be carried out. This is accomplished by the menu item 'Segmentation -> Save As Mesh…', following prompts, browsing to the location where the reconstructed model is to be stored, and giving a name in the format 'path\*.stl' for the file that represents the reconstructed 3D model (where 'path' is the complete path, e.g., C:\Users\PC\Desktop\*.stl, and * is any file name).

Now, a 3D reconstruction of the left kidney for the visible human male is over. Similar process may be followed to reconstruct the right kidney of the visible human male, and the left and right kidneys of the visible human female.

*Using MeshLab to Reduce the Total Number of Faces Describing the 3D Model*

The 3D model of kidney obtained through the use of ITK-SNAP typically is of very large size and typically is described by a very large number of surface triangles. MeshLab could be very helpful in reducing the total number of surface triangles that are needed to describe the 3D model satisfactorily. It also serves as a tool to smoothen the reconstructed 3D geometry; after using smoothing features provided by MeshLab, it may be necessary to scale the reconstructed 3D models to the correct dimensions, if the original dimensions are to be strictly retained. MeshLab can also improve the triangle quality of surface triangles of the 3D model. It can also reduce the file size.

The 3D models of kidney, after undergoing processing with MeshLab, are shown in the next section.

**Results**

Reconstructed left kidney of the male, after undergoing processing through MeshLab, is shown in Figure 5. Similarly, reconstructed right kidney of the male, after undergoing processing through MeshLab, is shown in Figure 6. Reconstructed left kidney of the female is shown in Figure 7. Reconstructed right kidney of the female is shown in Figure 8. All the four 3D models are made up of 1000 surface triangles. While obtaining these four models, job of MeshLab is to smoothen the 3D models reconstructed through ITK-SNAP and to reduce the total number of surface triangles to 1000.

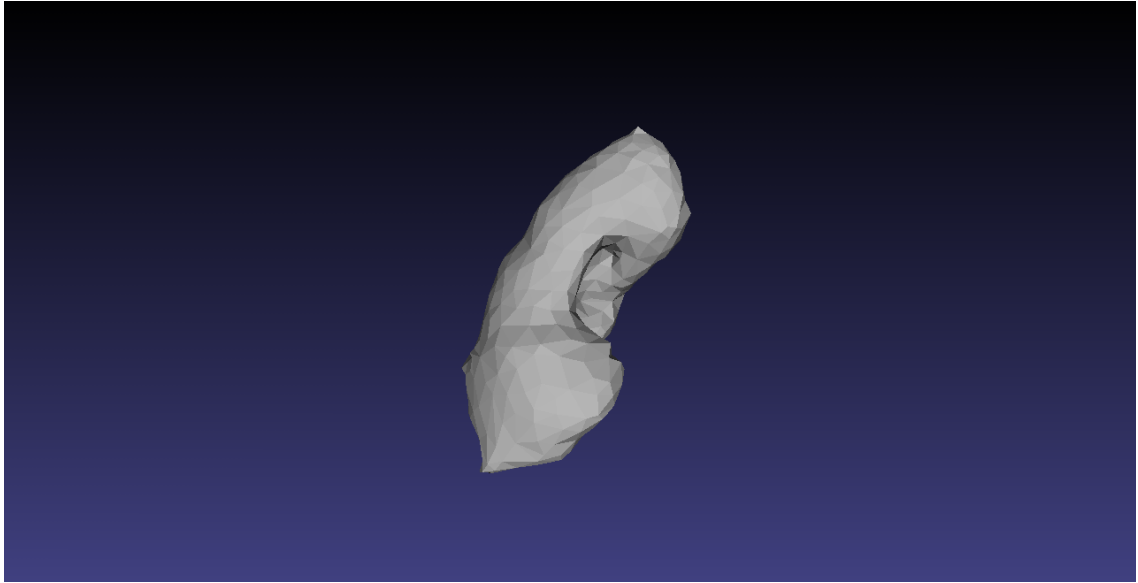

*Figure 5.* Reconstructed left kidney of male

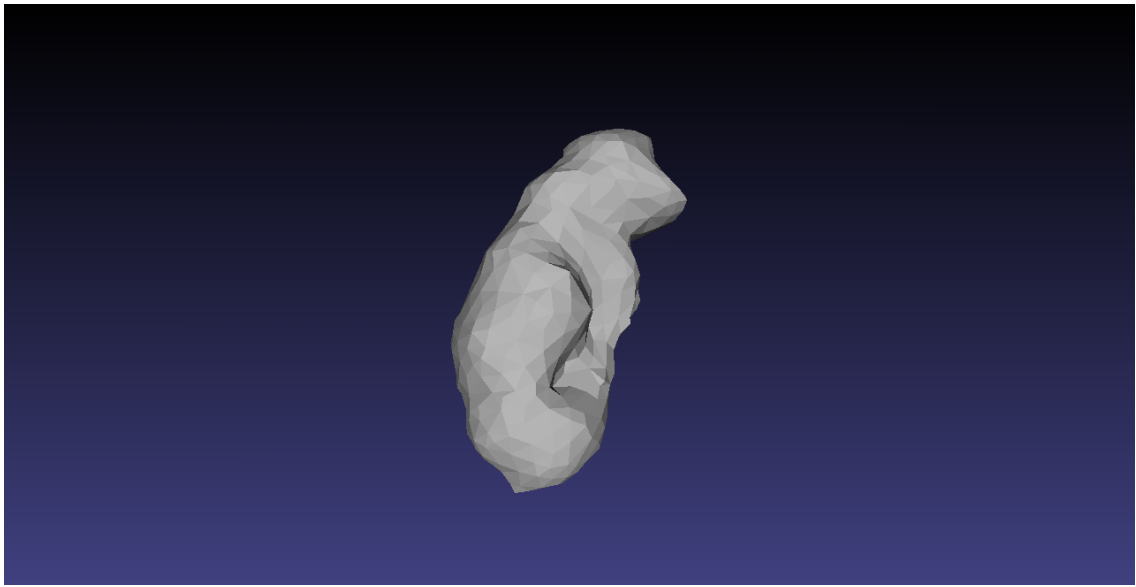

*Figure 6.* Reconstructed right kidney of male

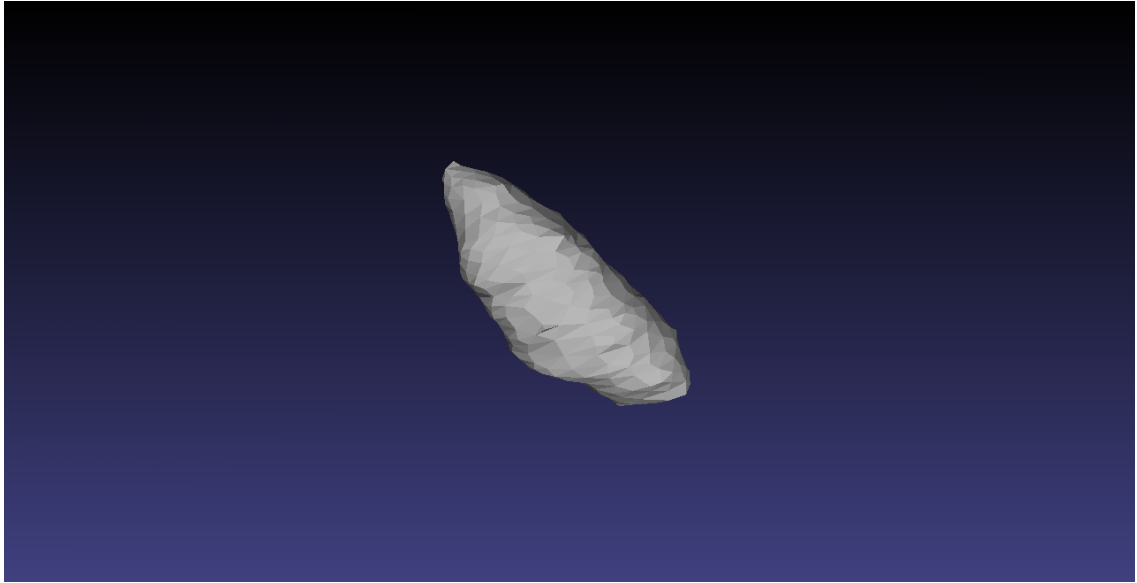

*Figure 7.  Reconstructed left kidney of female*

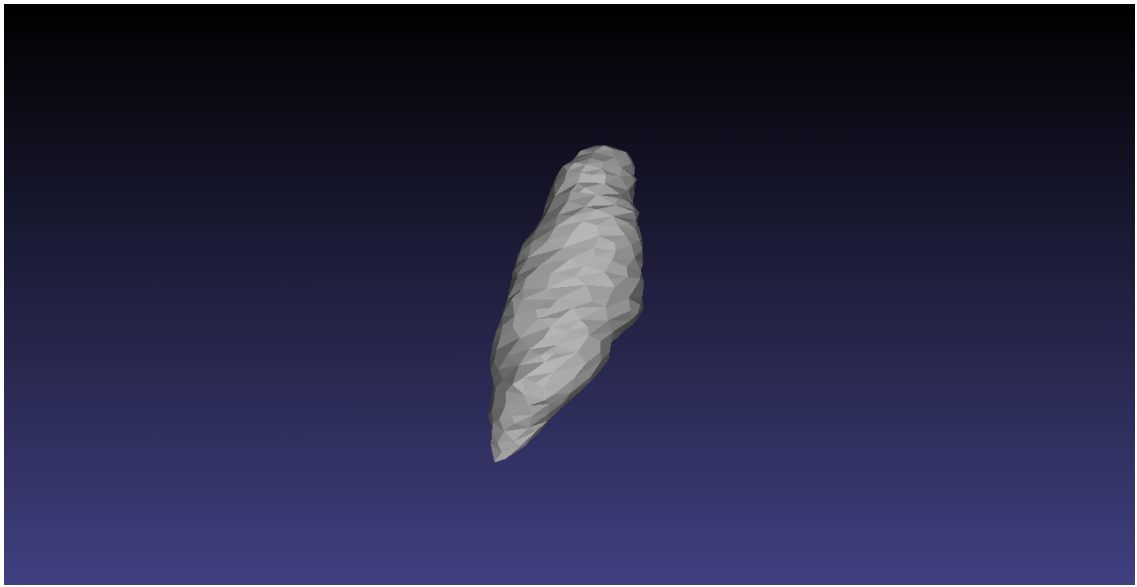

*Figure 8.  Reconstructed right kidney of female*

**Discussion**

In this work, 3D model of human kidney is extracted from CT-scan images from the VHD, using free software packages. The free software packages used are: 1) ImageJ 2) ITK-SNAP 3) MeshLab. The organs reconstructed are: 1) left kidney of visible human male 2) right kidney of visible human male 3) left kidney of visible human female 4) right kidney of visible human female. All the four models are in 'stl' format.

Use of free software packages together with images that may be obtained for free, as has been done in the present work, makes it possible to obtain the geometry of a representative human kidney, completely for free. Buying a 3D model of a human kidney, or using a commercial software package to extract 3D models from image sequences, cost (sometimes significant amount of) money. Also, in the present approach, user can control how finely the geometry should be described (using the free software package MeshLab). Since MeshLab can improve the quality of the surface mesh that describes a reconstructed 3D model, the reconstructed 3D model that has undergone processing with MeshLab can be used in a finite element analysis after converting the surface model to a solid model using software packages like Rhinoceros. Also, the method used to extract the geometry of a kidney, as illustrated in the present work, may possibly be used to extract other whole biological organs from VHD.

It may be noted that the method given here to obtain the 3D models of human kidney need not be followed rigidly. It is good to read the documentation for the software packages used here, and one can experiment with the various options provided by the software packages instead of rigidly following the method illustrated in this work. For example, instead of tracing the boundary of the kidney in each of the images through the mouse pointer, the 'Paintbrush tool' provided by ITK-SNAP can be tried out to carry out the segmentation; ITK-SNAP also provides a tool that can do semi-automatic segmentation.

As to the limitations, present work uses only CT-scan images. Although these are found to be sufficient to obtain the geometry of a whole kidney, whenever the reconstructed geometry should include the finer details of the kidney, or whenever some other organ is to be extracted from VHD, there is a possibility that other types of images (e.g., MRI images) are more suited in some cases. Also, multiple software packages need to be downloaded, installed and used here.

Future work is to extract other biological organs from VHD, using free software packages. Aim is to reconstruct biological organs with inner details (not obtaining just the outer surface of the organs) and to use other types of images (like MRI, cryosection images) from the VHD if need be.

## Conclusion

It is possible to obtain the 3D surface model of a representative human kidney from CT-scan images from the VHD, using free software packages only; the free software packages needed are: ImageJ, ITK-SNAP, and MeshLab. The practice of extracting the geometry of a representative human kidney completely for free, as illustrated in the present work, could be a free alternative to the use of expensive commercial software packages or to the purchase of a digital model.

## Acknowledgements

Author is grateful to the Robotics Lab, Department of Mechanical Engineering & Centre for Product Design and Manufacturing, Indian Institute of Science, Bangalore, INDIA, for providing the necessary infrastructure to carry out this work.

Author acknowledges Prof. Ashitava Ghosal, Robotics Lab, Department of Mechanical Engineering & Centre for Product Design and Manufacturing, Indian Institute of Science, Bangalore, INDIA, for providing the CT-scan images from the Visible Human Data Set (VHD).

Author acknowledges National Library of Medicine (NLM) and Visible Human Project (VHP) for providing free access to the Visible Human Data Set (VHD) to Prof. Ashitava Ghosal.

Visible Human Data Set (VHD) is an anatomical data set developed under a contract from the National Library of Medicine (NLM) by the Departments of Cellular and Structural Biology, and Radiology, University of Colorado School of Medicine.